\documentclass[a4paper,twoside,12pt]{article}

\usepackage{PRIMEarxiv}

\usepackage[utf8]{inputenc} 
\usepackage[T1]{fontenc}    
\usepackage{hyperref}       
\usepackage{url}            
\usepackage{booktabs}       
\usepackage{adjustbox}
\usepackage{amsfonts}       
\usepackage{nicefrac}       
\usepackage{microtype}      
\usepackage{lipsum}
\usepackage{fancyhdr}       
\usepackage{graphicx}       
\graphicspath{{media/}}     
\usepackage{adjustbox}%
\usepackage{subcaption}
\usepackage{scalefnt}
\usepackage{color}

\title{Human Preferences for Constructive Interactions in Language Model Alignment
}

\author{
  Yara Kyrychenko \\
  Department of Psychology \\
  University of Cambridge \\
  Cambridge, UK\\
  \texttt{yk408@cam.ac.uk} \\
    \And
  Jon Roozenbeek\\
  Department of War Studies \\
  King's College London \\
  London, UK\\
    \texttt{jon.roozenbeek@kcl.ac.uk}
    \And
  Brandon Davidson \\
  Department of Psychology \\
  University of Cambridge \\
  Cambridge, UK\\
  \texttt{bd440@cam.ac.uk} \\
   \And
  Sander van der Linden\\
  Department of Psychology \\
  University of Cambridge \\
  Cambridge, UK\\
  \texttt{sv395@cam.ac.uk}
   \And
  Ramit Debnath \\
  Collective Intelligence \& Design Group \\
  University of Cambridge \\
  Cambridge, UK\\
  \texttt{rd545@cam.ac.uk} \\
}

\begin{document}
\maketitle

\begin{abstract}
As large language models (LLMs) enter the mainstream, aligning them to foster constructive dialogue rather than exacerbate societal divisions is critical. Using an individualized and multicultural alignment dataset of over 7,500 conversations of individuals from 74 countries engaging with 21 LLMs, we examined how linguistic attributes linked to constructive interactions are reflected in human preference data used for training AI. We found that users consistently preferred well-reasoned and nuanced responses while rejecting those high in personal storytelling. However, users who believed that AI should reflect their values tended to place less preference on reasoning in LLM responses and more on curiosity. Encouragingly, we observed that users could set the tone for how constructive their conversation would be, as LLMs mirrored linguistic attributes, including toxicity, in user queries.
\end{abstract}

Large language models (LLMs) such as ChatGPT have become an important component of our daily lives \cite{statista2024}, making it crucial to understand what values these models reflect. Previous research shows that LLMs can exhibit social, racial, and religious biases similar to humans \cite{hu2024,bordia_identifying_2019,abid2021persistent}, raising concerns about their potential to worsen societal rifts. To address this, recent efforts have focused on aligning LLMs with humans using preference data--pairs of user queries and LLM response options, with one response preferred over the rest (e.g., ref. \cite{bai2022training,askell2021general}). However, the preferences of humans are not immune to their biases, which could lead LLMs to reproduce unintended values \cite{sharma2024towards,li2024dissecting}. Therefore, it is critical to investigate whether we are aligning our AI to bridge divides or unintentionally deepen them. 

Here, we examine how language attributes that foster constructive and unconstructive interactions are reflected in human preference data. We focus on bridging attributes\footnote{See Table \ref{tab:defs} for definitions from the Perspective API \cite{jigsaw2024} used here.}, such as respect and compassion, that promote constructive dialogue \cite{jigsawbridging, bridgingsystems, bao2021llms} and toxicity, which is linked to reduced civility online \cite{avalle2024nature}. We also conduct a moderation analysis to see how user factors such as gender, age, ethnicity, and emphasis on AI value alignment influence which language attributes are associated with a higher score. We find that reasoning and nuance are the strongest predictors of higher ratings, whereas more personable LLM responses (e.g., exhibiting personal storytelling or curiosity) generally score lower. Although there is broad agreement in the degree of preference for affinity, compassion, and respect, users who believe that AI should reflect their values tend to rate reasoning lower and curiosity higher. Finally, we observe that users can steer how constructive their conversation would be through prompting and toxic user prompts are associated with more toxic but also more compassionate responses. We also observe that users who prompt with toxicity are more likely to give higher ratings to toxic LLM responses, highlighting the risks of personalized LLM alignment. These insights contribute to the ongoing efforts in AI alignment and refine our understanding of human-AI interaction.

\begin{figure}
    \centering

    \vspace{-0.5cm}

    \begin{subfigure}[t]{.55\textwidth}
        \centering
        \caption{\textbf{a} LLM response attributes predicting the highest score}
        \includegraphics[width=\linewidth,trim={0 0 0 0.5cm}, clip]{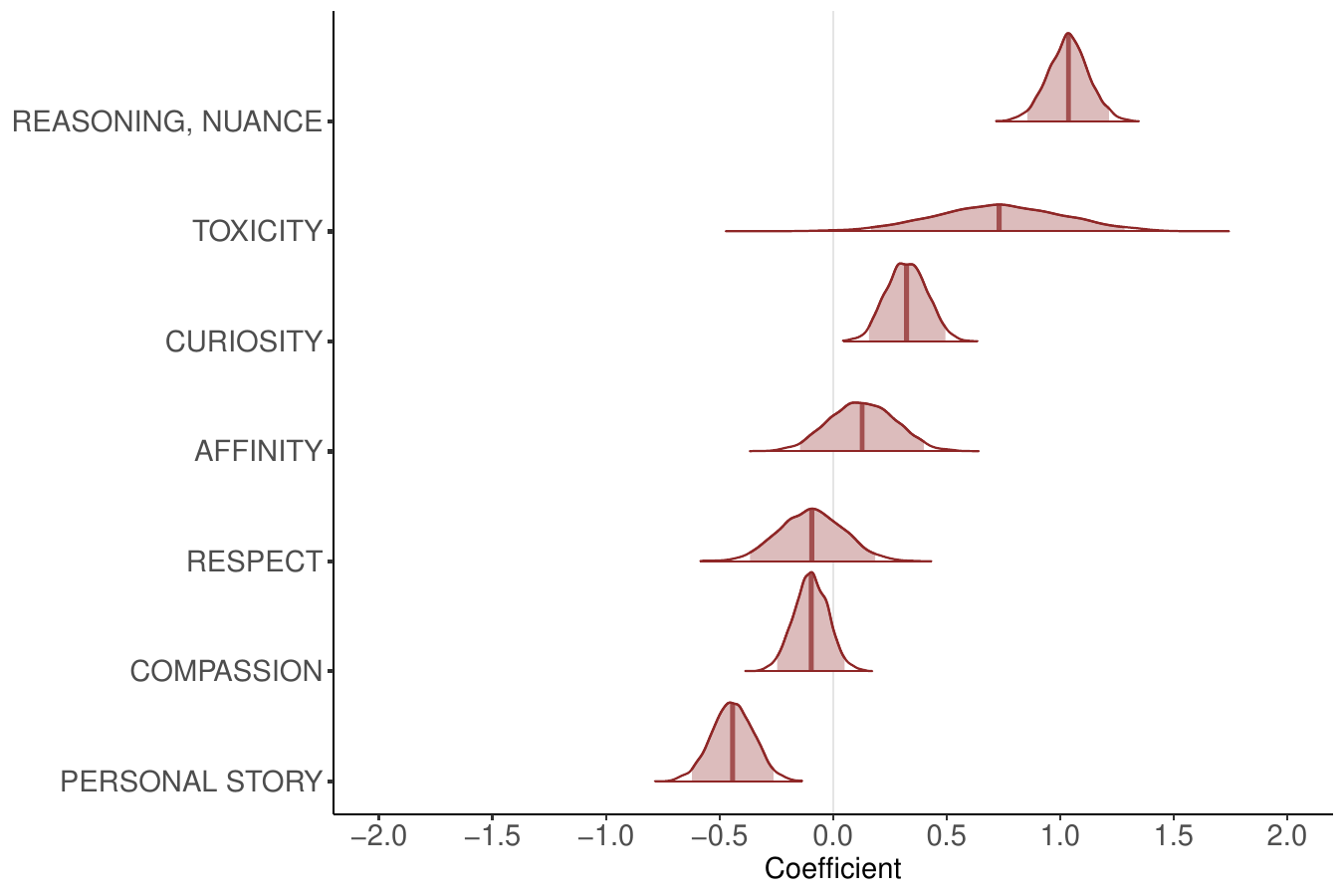}
        \label{fig:unguided}
    \end{subfigure}%
    \begin{subfigure}[t]{.411\textwidth}
        \centering
        \caption{\textbf{b} LLM response attributes predicting log-score}
        \includegraphics[width=\linewidth,trim={5.8cm 0 0 0.5cm}, clip]{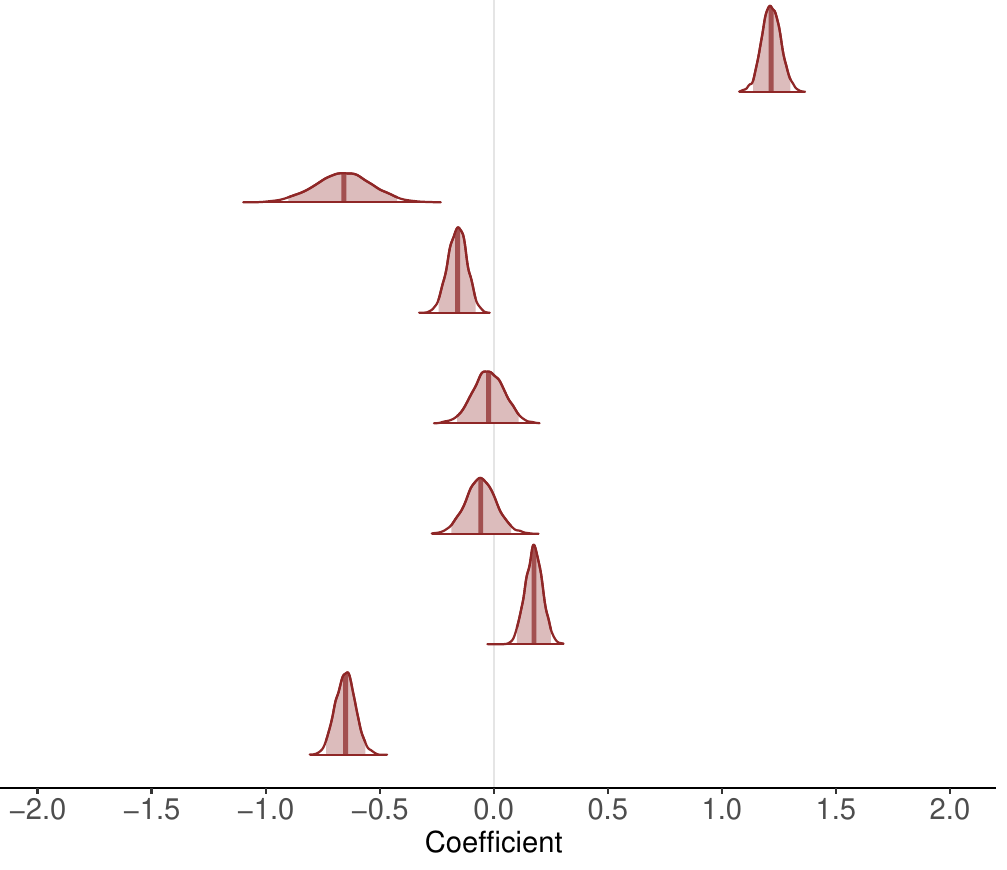}
        \label{fig:value}
    \end{subfigure}%

    \begin{subfigure}[t]{\textwidth}
        \centering
        \caption{\textbf{c} Human prompt attributes predicting LLM response attributes}
        \includegraphics[width=\linewidth,trim={0 0 0 0}, clip]{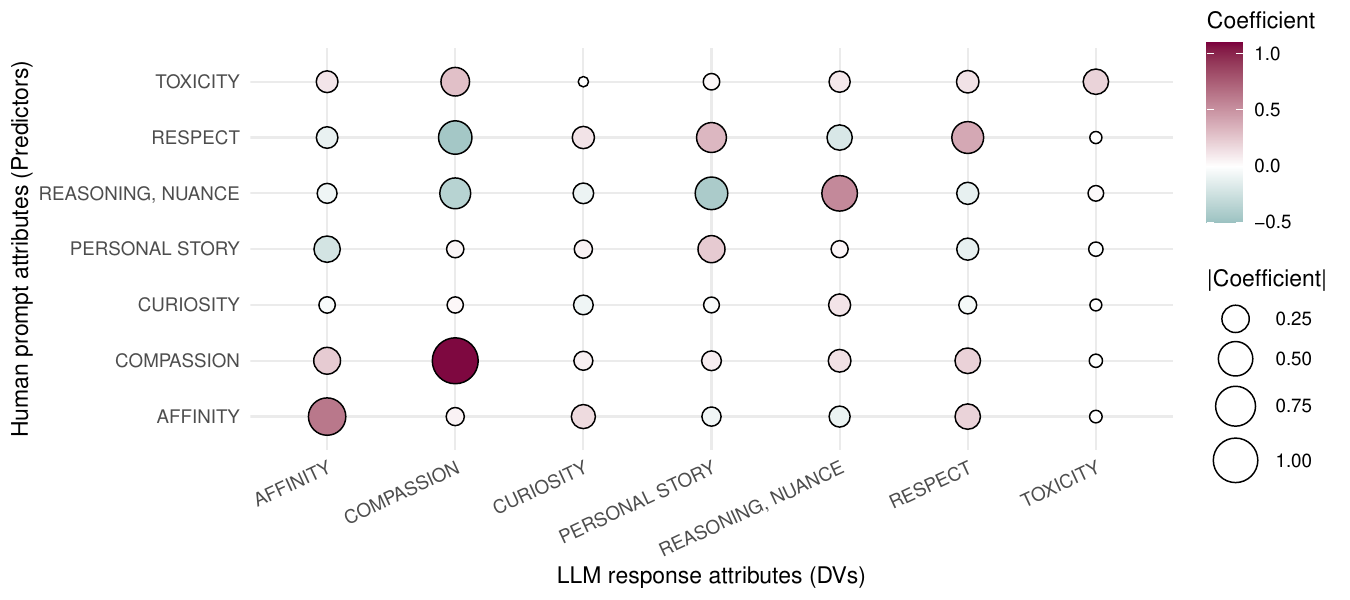}
        \label{fig:value}
    \end{subfigure}%
    
    \vspace{-0.5cm}

    \caption{Posterior distributions of coefficients from the mixed-effects Bayesian logistic and linear regressions, with 95\% credible intervals, for the highest score (a) and score (b) of LLM responses. (c) Coefficients of mixed-effects Bayesian linear regressions predicting LLM response attributes based on human query attributes (higher estimates indicate a higher likelihood of response exhibiting the attribute).}

    \label{fig:mainfig}
\end{figure}

\section*{Results}

Using Bayesian modeling\cite{burkner2017brms}, we examined the human preferences for constructive LLM responses encoded in an individualized and multicultural alignment dataset comprising over 7,500 human-LLM conversations \cite{prism2024}. We performed mixed-effects Bayesian logistic regressions, predicting which response got the highest score from human participants out of two to four options, given the probability of each of the bridging attributes (with reasoning and nuance multiplied due to collinearity) and toxicity, as classified by the Google Jigsaw Perspective API \cite{jigsaw2024}. Achieving the highest score is important as it determines the subsequent conversational model and the final preference choice for fine-tuning (e.g., \cite{bai2022training,rafailov2023dpo}). We also performed mixed-effect Bayesian linear regressions with the user score of each response as the dependent variable and the same predictors as above to assess the more fine-grained human preferences. We then ran a mixed-effects Bayesian linear regression predicting score to test for the interaction effects of age, gender, ethnicity, and AI value alignment preference. Finally, we conducted mixed-effects Bayesian linear regressions predicting how bridging attributes in model responses depend on the levels of each attribute in user prompts. 

\begin{table*}[t]
\caption{Descriptions of bridging attributes and toxicity by Google Jigsaw Perspective API team \cite{jigsaw2024}.}

\centering

\begin{adjustbox}{width=1\textwidth}
\begin{tabular}{ll}
\toprule
\textbf{Category} & \textbf{Description} \\ 
\midrule
AFFINITY & References shared interests, motivations or outlooks between the comment author and another individual, group or entity. \\  
COMPASSION & Identifies with or shows concern, empathy, or support for the feelings/emotions of others. \\  
CURIOSITY & Attempts to clarify or ask follow-up questions to better understand another person or idea. \\ 
NUANCE & Incorporates multiple points of view in an attempt to provide a full picture or contribute useful detail and/or context. \\ 
PERSONAL STORY & Includes a personal experience or story as a source of support for the statements made in the comment. \\  
REASONING & Makes specific or well-reasoned points to provide a fuller understanding of the topic without disrespect or provocation. \\  
RESPECT & Shows deference or appreciation to others, or acknowledges the validity of another person. \\ 
TOXICITY & A rude, disrespectful, or unreasonable comment that is likely to make people leave a discussion. \\ \hline
\end{tabular}
\end{adjustbox}
\label{tab:defs}

\end{table*}

\paragraph{Analysis 1a.} We found that reasoning and nuance were the strongest predictors of the highest response score (see Figure \ref{fig:mainfig}a and Table \ref{tab:mainfindings}a).  Controlling for all other variables, a 10\% increase in the probability of an LLM's output exhibiting reasoning and nuanced language increased its odds of achieving the highest score by 18.18\% ($\beta=1.036$, 95\% Credible Interval (CI) = $[0.888,1.185]$, Bayes Factor (BF) = 4,000;  $(\exp(\beta)-1)*10\%=18.18\%$), with an 11.51\% increase in unguided conversations, 26.84\% in value-guided conversations, and 32.76\% in controversy-guided conversations. Curiosity also increased the probability of response scoring the highest by 3.83\% ($\beta=0.324$, 95\% CI = $[0.181,0.469 ]$, BF = 4,000) in all conversations and 6.62\% ($\beta=0.508$, 95\% CI = $[0.286,0.734]$, BF = 3,999) in unguided conversations, per 10\% increase. On the other hand, a 10\% increase in personal storytelling was associated with a 3.57\% decrease in the likelihood of scoring the highest across all conversations ($\beta=-0.441$, 95\% CI = $[-0.595, -0.293]$, BF = 4,000). Surprisingly, toxicity was significantly associated with response scoring the highest across all conversation types: controlling for all else, a 10\% increase in toxicity increased the odds of the response scoring the highest by 10.73\% ($\beta=0.729$, 95\% CI = $[0.259,1.195]$, BF = 141.857). This was not the case for unguided and value-guided conversations, but in controversial conversations, a 10\% increase in toxicity contributed to a 14.94\% increase ($\beta=0.914$, 95\% CI = $[0.268, 1.555]$, BF = 124).

\paragraph{Analysis 1b.} Reasoning and nuance were the strongest predictors of higher response scores (see Figure \ref{fig:mainfig}b and Table \ref{tab:mainfindings}b). Controlling for all other variables, a 10\% increase in the probability of reasoning and nuance increased the score by 23.77\% overall ($\beta=1.217$, 95\% CI = $[1.150,1.286]$, BF = 4,000;  $(\exp(\beta)-1)*10=23.77$), increasing by 17.68\%, 24.94\%, and 38.11\% for unguided, value-guided and controversy-guided conversations, respectively. Compassion also predicted higher scores, albeit to a lesser extent, with a 1.92\% increase ($\beta=0.176$, 95\% CI = $[0.112,0.239]$, BF = 3999) across all conversations for a 10\% increase. Meanwhile, curiosity and personal storytelling were associated with lower scores: an increase of 10\% was associated with a 1.47\% ($\beta=-0.159$, 95\% CI = $[-0.229,-0.092]$, BF = 4,000) and 4.78\% ($\beta=-0.650$, 95\% CI = $[-0.723, -0.577]$, BF = 4,000) decrease in score across all conversations, respectively.  A 10\% increase in toxicity predicted a 4.83\% decrease in score across all conversations ($\beta=-0.659$, 95\% CI = $[-0.865, -0.458 ]$, BF = 4,000). However, the interaction of model response toxicity and user query toxicity was positive, indicating that users who prompted with more toxic content also rated toxic LLM responses higher. 

\paragraph{Analysis 2.} For the moderation analysis, we found that users under 45 years old rated toxic or personal-story responses lower compared to those over 45, controlling for all other interaction terms (see Table \ref{tab:moderation}). We also found that users who believed that AI should reflect their values and cultural perspectives rated responses demonstrating curiosity higher and responses with reasoning and nuance lower, as compared to users who thought value alignment less important. 

\paragraph{Analysis 3.} We observed that LLMs tend to echo the linguistic attributes from user queries in their responses (see Figure \ref{fig:mainfig}c and Table \ref{tab:users}). Specifically, user prompts characterized by respect, reasoning, nuance, personal storytelling, compassion, and affinity prompted similar responses from the models. Curiosity in user prompts led to more reasoning and nuance in model responses, while nuanced reasoning in prompts was linked to lower compassion, respect, and personal storytelling in LLM output. Notably, respect in human queries was correlated with less compassion and more personal storytelling. Finally, toxic human queries were associated with increased model compassion and were the largest factor predicting increased toxicity in LLM responses.

\section*{Discussion}

As AI systems become more integrated into society, it is crucial to ensure they promote constructive engagement rather than deepening societal divisions \cite{shen2024towards}. Values like compassion and respect tend to foster positive interactions across divides \cite{bridgingsystems,saltz2024re}, yet our findings reveal that these values might not be well-represented in AI training data. 

Analyzing a multicultural AI-alignment \cite{prism2024} dataset of human conversations with 21 different LLMs in English, we find that people appear to prefer well-reasoned and nuanced LLM responses more than compassionate, curious, or respectful answers, while responses containing personal storytelling are dispreferred. Of course, attributes like compassion are contextual and might be desired only in certain situations \cite{saltz2024re,kyrychenko2025c3ai}, yet compassion does increase scores on average. 

We also observe some individual differences in these preferences, reiterating the difficulties of aggregate human-LLM alignment \cite{kirk2024benefits,conitzer2024social}. Since we find that users who prompt with toxicity tend to rate toxic LLM responses higher, LLMs personalized to an individual might create feedback loops of negative content \cite{wolf2023fundamental}. However, some preferences, such as those for affinity, compassion, and respect, seem relatively consistent across gender, age, ethnicity, and views on AI value alignment. Finally, the fact that we observe different outcomes for the binary preference versus response scores (especially for toxicity and curiosity) suggests that the current methodology of pairwise preference fine-tuning might be missing important details by simplifying human feedback \cite{wolf2023fundamental,rodemann2025statistical}. 

Our analysis of how user query attributes predict LLM response attributes also indicates that LLMs are prone to echoing user conversational characteristics even when they are toxic, similar to previous work demonstrating that LLMs tend to match human beliefs even when they are fake \cite{sharma2024towards}. Nonetheless, this gives some degree of control over the constructiveness of human-LLM conversations back to the users as, for instance, compassionate queries are met with more compassion. Remarkably, toxic queries are also more likely to receive a compassionate reply.

\section*{Methods}

We analyzed the PRISM dataset of human-LLM conversations of 1,394 individuals from 74 countries and 21 different language models \cite{prism2024}. Participants gave their demographic attributes and engaged in several conversations with LLMs, where they were first asked to choose the type of conversation they wanted to have: an unguided conversation about a random topic, a controversy-guided conversation about topics the participant believes are controversial, or a value-guided conversation about topics that are important to the participant or represent their values. Participants were instructed to begin the conversation with a question, request, or statement. They then rated each of the response options from Terrible to Perfect using a 100-point slider, and the model with the highest score was chosen for subsequent conversational turns. This resulted in a total of 7,960 English-language human-LLM conversations (3,093 unguided, 2,427 controversy-guided, and 2,440 values-guided). 

Next, we used the Google Jigsaw Perspective API \cite{jigsaw2024} on the first user prompt and the LLM responses in each conversation to obtain the probability that a response exhibited bridging attributes (affinity, compassion, curiosity, reasoning and nuance, a personal story, and respect) and toxicity (see Table \ref{tab:defs} for definitions). We note that due to collinearity, we collapsed the reasoning and nuance categories into a single variable.

We ran mixed-effects Bayesian regressions, with random effects of conversations nested within users and of LLM names, to assess whether the language attributes predict (\textbf{Analysis 1a}) whether the response was scored the highest of all response options (0 or 1), and (\textbf{Analysis 1b}) the response score (1-100). We fit one model with all the data and three separate models for each type of conversation. We also conducted a moderation analysis (\textbf{Analysis 2}) \cite{homan2023intersectionality} using a mixed-effects Bayesian regression predicting score to see if our results vary by age (under vs. over 45), gender (male vs. non-male), ethnicity (white vs. non-white), and value alignment (i.e., rating over vs. under 50 on a 100-point scale for the statement \textit{It is important that an AI language model reflects my values or cultural perspectives}). In all regressions, score was log-transformed. Finally, to investigate how users are shaping LLM responses, we fit a mixed-effects Bayesian regression predicting how model response attributes depend on the levels of each of the linguistic attributes in user prompts (\textbf{Analysis 3}). We used a Bayes Factor (BF) of 100 as a cut-off point for significance, considered definitive (or extreme) evidence for the alternative hypothesis \cite{stefan2019bayes}, and performed one-sided hypothesis tests.

\section*{Code and Data Availability}
All code and data needed to reproduce the analyses in this paper is available on \href{https://osf.io/7xtyc/?view_only=6070f79da7904c00a86e5b4388244d5e}{OSF}. 

\section*{Acknowledgements}
Y.K. is supported by Gates Cambridge Trust (grant OPP1144 from the Bill \& Melinda Gates Foundation) and the Alan Turing Institute's Enrichment Scheme. R.D. acknowledges the support from CHRG AHSS grants, ai@cam AIDEAS grant, and UKRI Responsible AI UK grants. The authors thank Julian Wykowski for helpful discussions.

\section*{Author Contributions}
Y.K. conceptualized the study and analyzed the data. J.R., S.v.d.L., and R.D. helped with the conceptualization and provided feedback. J.R., R.D., and Y.K. led the write-up. B.D assisted with data analysis and write-up. 

\section*{Competing Interests}
The authors declare no competing interests.

\bibliographystyle{unsrt}  
\bibliography{references}  

\newpage

\appendix

\section*{Supplementary Information}

\subsection*{Supplementary Methods}
We found that the Bayesian regression model fit on all data with random effects of both (1) conversations nested within users and (2) LLMs had a higher expected log predictive density than the model without random effects of LLMs (by 504.9 with SE 30.5). Therefore, we opted to fit all models with both random effects. We chose very weakly informative priors -- normal distribution with mean zero and standard deviation of $10^6$ \cite{aki2020rank} -- and used 4 Markov Chains, 1000 iterations for warmup, and 1000 after warmup in Analyses 1a and 1b and 3,000 after warmup for Analyses 2 and 3 to get higher resolution. The score was always logarithmically transformed. 

All model coefficients and diagnostic values are under the results section of our OSF. For the models predicting the highest score (all types of data and conversations) and score (all types of data and unguided conversations only), the $\hat{R}$ values of 1 and high Bulk and Tail ESS indicated that the chains converged and had high resolution. For the model predicting score of value-guided and controversial conversations, all estimates of effects had an $\hat{R}$ close to 1 and good ESS ($>400$), other than the random effect of user-id:conversation-id. We, therefore, fit additional models with only user ID as the random effect as a robustness check, which produced similar results.

We originally intended to use the full set of bridging and toxicity attributes (e.g., severe toxicity or identity attack) as classified by the Perspective API but observed high collinearity as measured by the variance inflation factors (VIFs) for toxicity, reasoning, and nuance. Given that the toxicity attributes other than toxicity itself had very low values and were highly inter-correlated, we did not include those in our regressions. For reasoning and nuance, we first used the de-trended versions of each variable in a new model to estimate how much their independent components predicted the highest score, finding roughly equal coefficients for both. Reasoning and nuance were also very highly correlated (around .9) and have similar conceptual definitions. Therefore, we combined them into one variable, which fixed the VIFs. For the moderation analysis, all \textit{Prefer not to say} responses were removed.
\newpage

\subsection*{Supplementary Figures}

\begin{figure}[tbh!]

    \centering

    \vspace{-1cm}

    \begin{subfigure}[b]{\textwidth}
        \centering
        \caption{\center LLM response attributes predicting the highest score}
    \end{subfigure}

    \begin{subfigure}[b]{.4\textwidth}
        \centering
        \caption{\textbf{a} Unguided conversations}
        \includegraphics[width=\linewidth,trim={0 0 0 0}, clip]{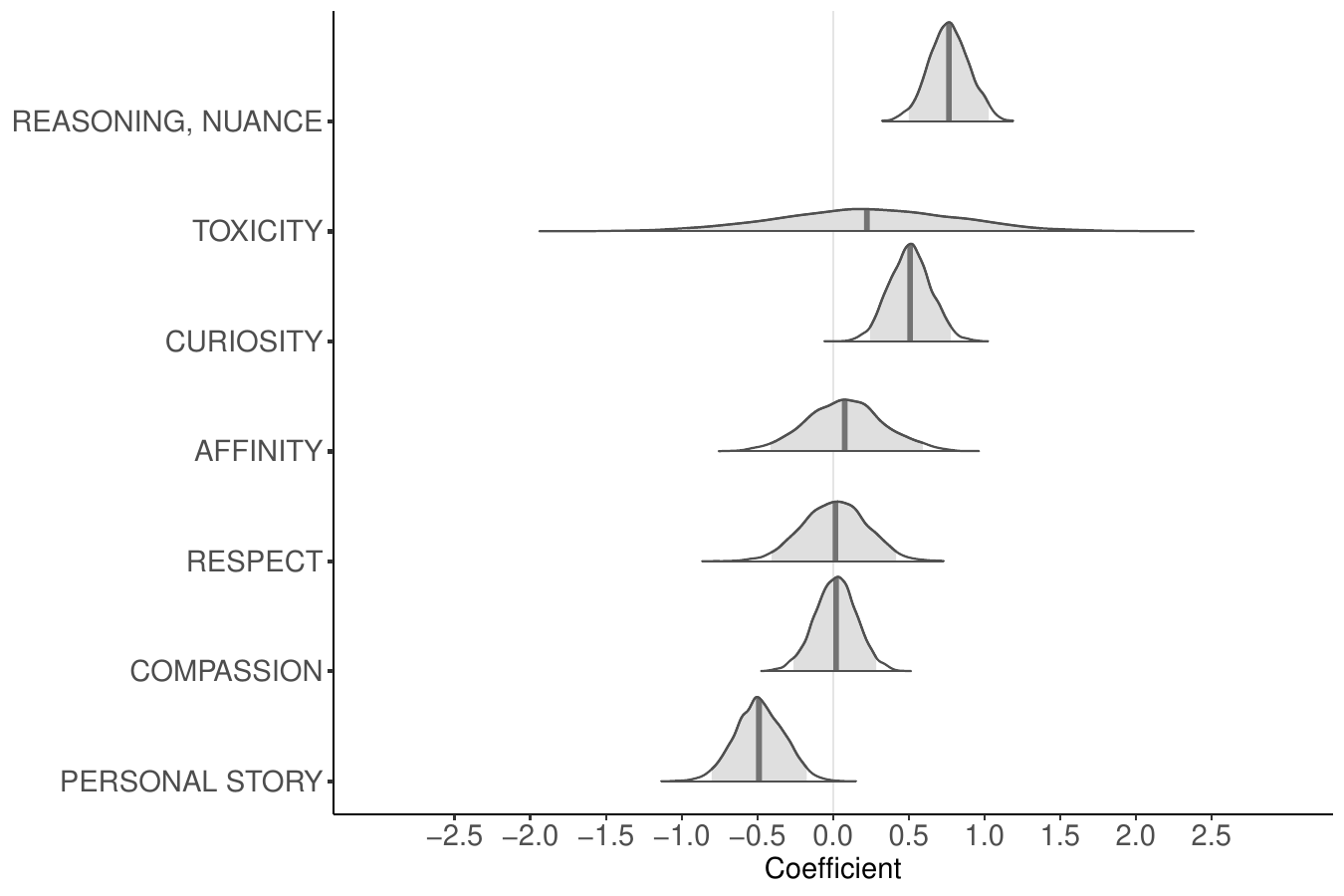}
        \label{fig:unguided}
    \end{subfigure}%
    \begin{subfigure}[b]{.3\textwidth}
        \centering
        \caption{\textbf{b} Value-guided conversations}
        \includegraphics[width=\linewidth,trim={5.8cm 0 0 0}, clip]{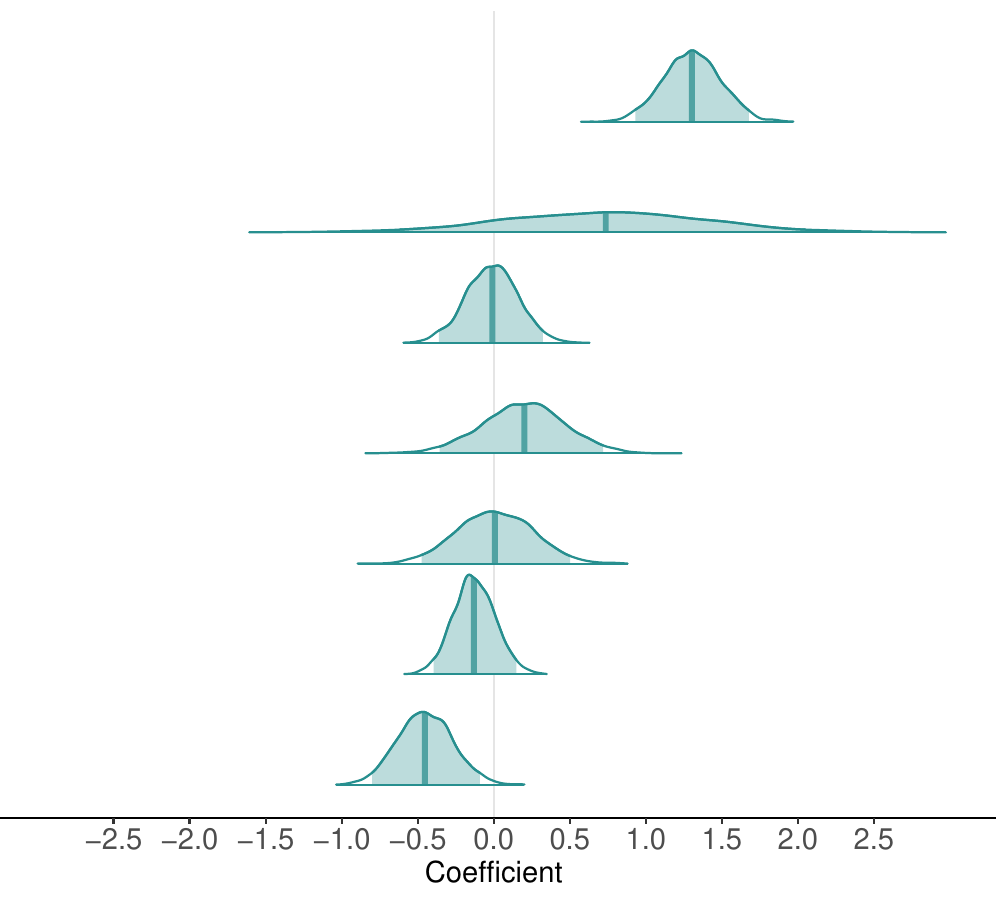}
        \label{fig:value}
    \end{subfigure}%
    \begin{subfigure}[b]{.3\textwidth}
        \centering
        \caption{\textbf{c} Controversy-guided conversations}
        \includegraphics[width=\linewidth,trim={5.8cm 0 0 0}, clip]{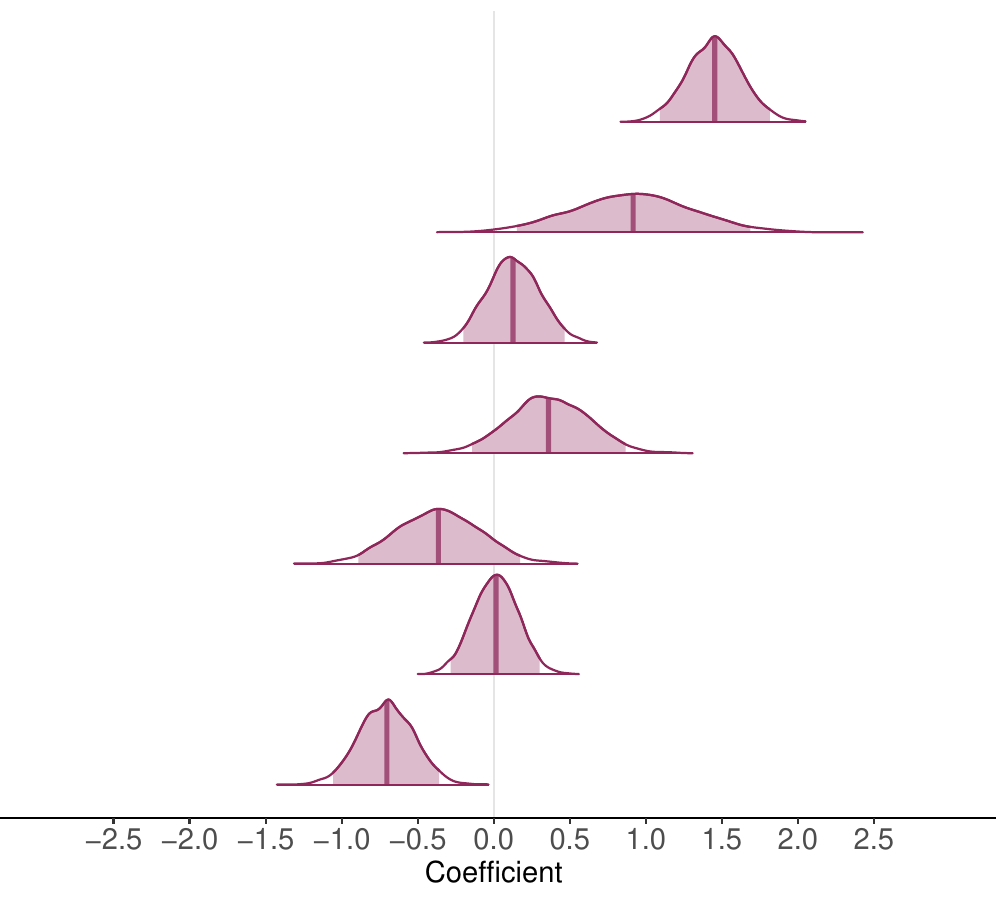}
        \label{fig:controversial}
    \end{subfigure}%

    \vspace{-.5cm}

    \begin{subfigure}[b]{\textwidth}
        \centering
        \caption{\center LLM response attributes predicting response score}
    \end{subfigure}

    \begin{subfigure}[b]{.4\textwidth}
        \centering
        \caption{\textbf{d} Unguided conversations}
        \includegraphics[width=\linewidth,trim={0 0 0 0}, clip]{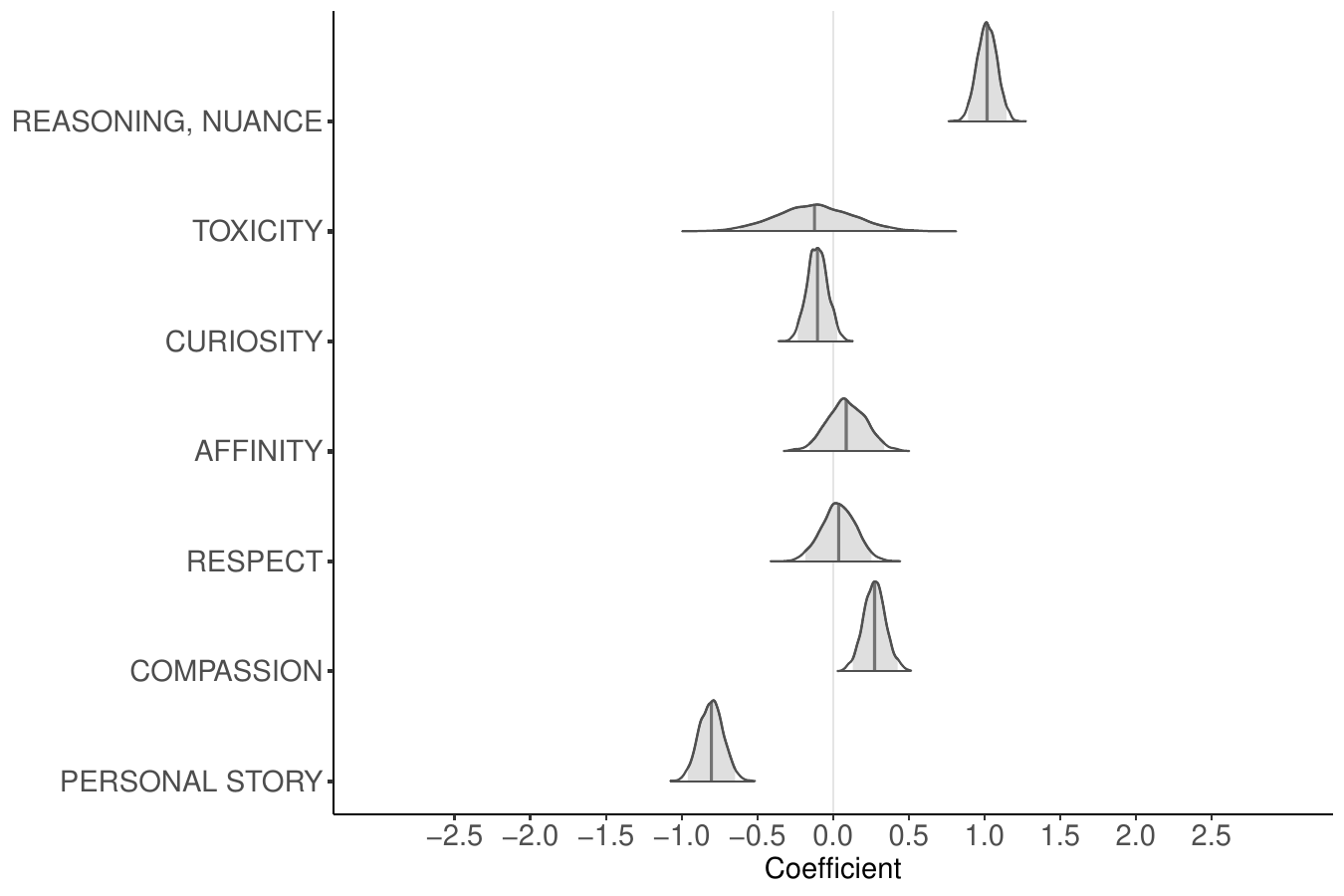}
        \label{fig:unguided}
    \end{subfigure}%
    \begin{subfigure}[b]{.3\textwidth}
        \centering
        \caption{\textbf{e} Value-guided conversations}
        \includegraphics[width=\linewidth,trim={5.8cm 0 0 0}, clip]{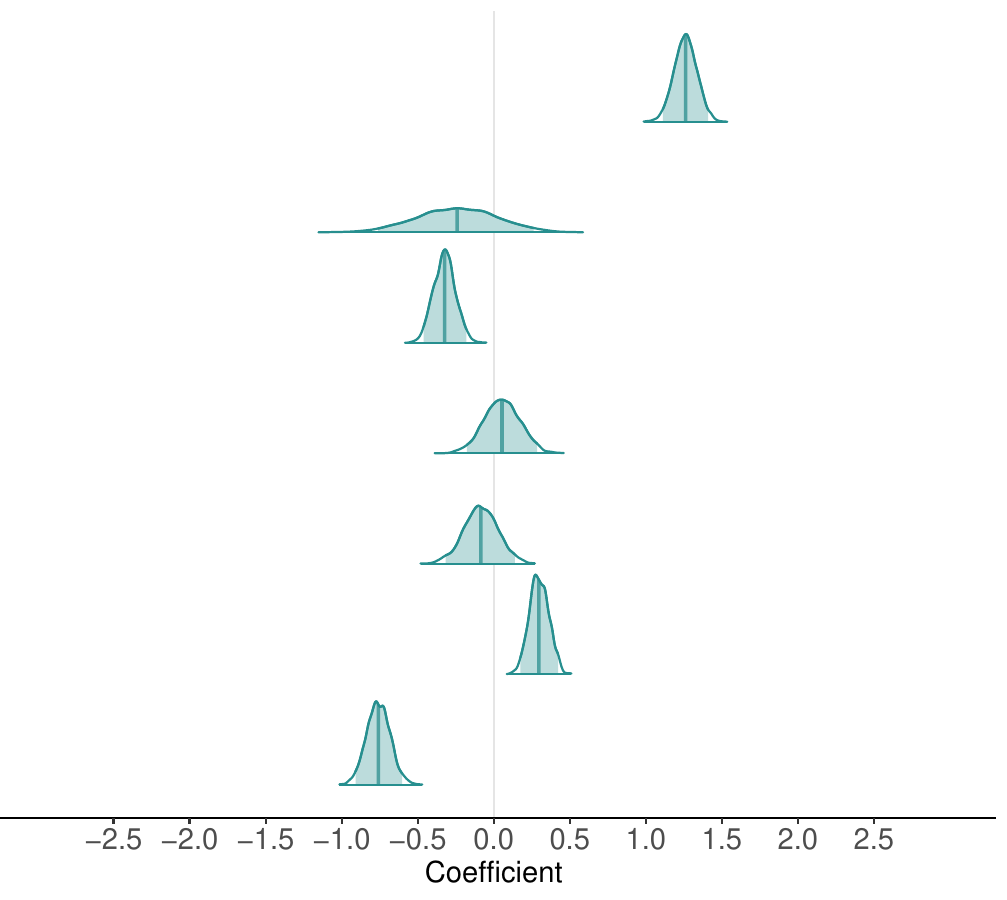}
        \label{fig:value}
    \end{subfigure}%
    \begin{subfigure}[b]{.3\textwidth}
        \centering
        \caption{\textbf{f} Controversy-guided conversations}
        \includegraphics[width=\linewidth,trim={5.8cm 0 0 0}, clip]{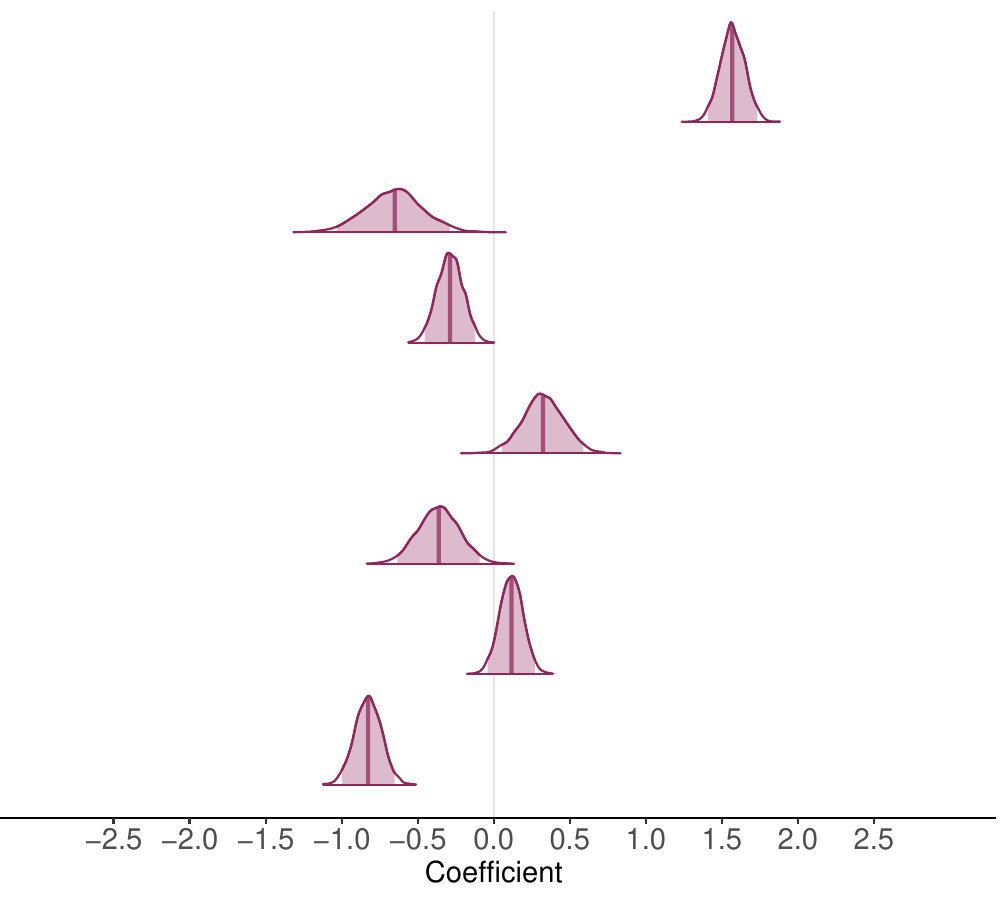}
        \label{fig:controversial}
    \end{subfigure}%

    \vspace{-.5cm}

    \begin{subfigure}[b]{\textwidth}
        \centering
        \caption{\center Probabilities of conversational attributes in LLM responses}
    \end{subfigure}

    \vspace{-.5cm}

    \begin{subfigure}[t]{.74\textwidth}
        \centering
        \caption{\textbf{g} Bridging attributes}
        \includegraphics[width=\linewidth, keepaspectratio,trim={0 0 0 1cm}, clip]{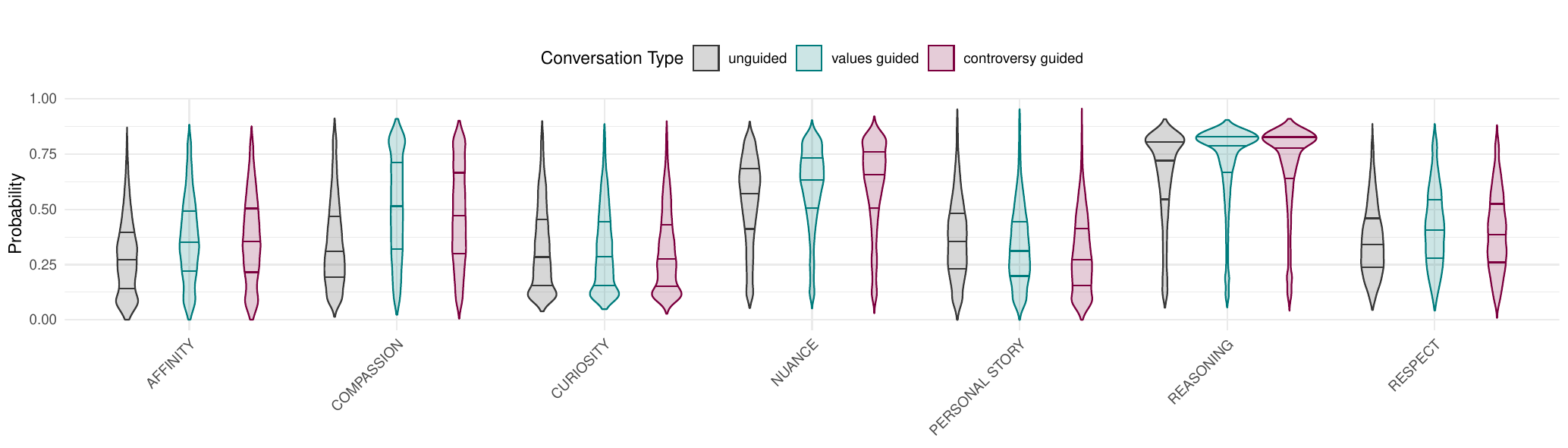}
    \end{subfigure} 
    \begin{subfigure}[t]{.225\textwidth}
        \centering
        \caption{\textbf{h} Toxicity}
        \includegraphics[width=\linewidth, keepaspectratio, trim={3cm 0 2cm 1cm}, clip]{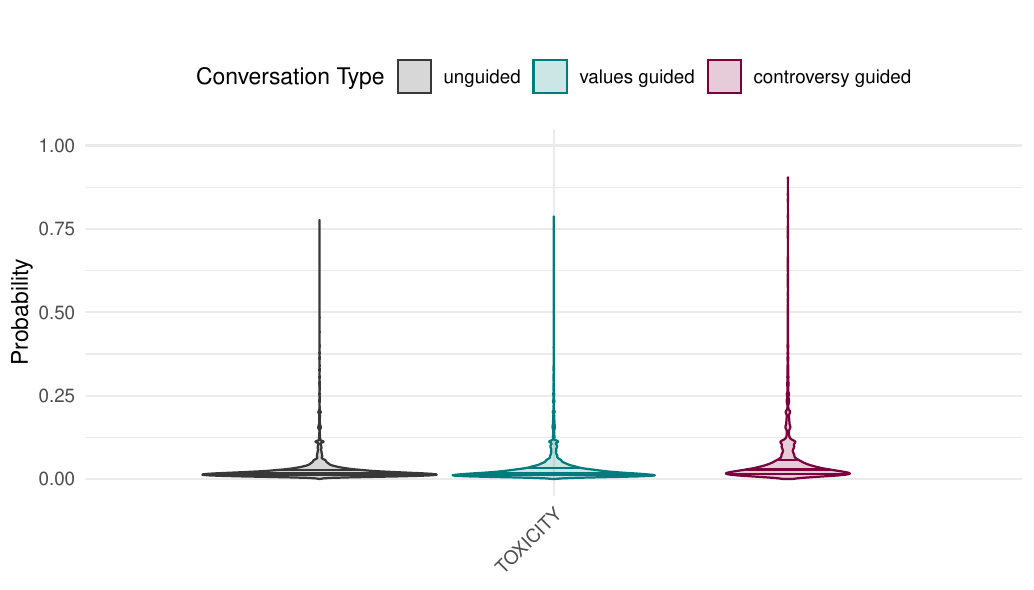}
    \end{subfigure}

    \caption{Posterior distributions of coefficients from the mixed-effects Bayesian logistic and linear regressions, with 95\% credible intervals, for user the highest score (panels a, b, and c) and score (panels d, e, and f) of LLM responses. Probabilities of bridging attributes (g) and toxicity (h) across three conversation categories.}

    \label{fig:fig_all}
\end{figure}

\newpage

\subsection*{Supplementary Tables}
\begin{table}[!htbp] 
\centering 
\caption{Results of hypothesis tests for models predicting the highest score (logistic regressions) and score (linear regressions; the score was log-transformed). In the ``Star'' column, $\ast$ indicates a BF $> 100$, our significance threshold indicating ``definitive'' or ``extreme'' evidence in favor of the alternative hypothesis, while $\dagger$ represents BF $> 10$. Note: $Inf$ should be interpreted as the maximum number of draws (4,000). BF calculated as Evidence Ratio.} 
\begin{adjustbox}{width=\textwidth}
\begin{tabular}{lcccccccc}
\\[-1.8ex] \hline \hline \\[-1.8ex] 
Hypothesis & Estimate & Est.Error & CI.Lower & CI.Upper & BF & Post.Prob & Star & Convo Type \\ 
\hline \\[-1.8ex] 
\multicolumn{9}{c}{(a) \textbf{Analysis 1a:} The highest score} \\ 
\hline
(REASONING\_NUANCE) \textgreater  0 & $1.036$ & $0.090$ & $0.888$ & $1.185$ & $Inf$ & $1$ & \textasteriskcentered  & all \\ 
(REASONING\_NUANCE) \textgreater  0 & $0.766$ & $0.133$ & $0.549$ & $0.992$ & $Inf$ & $1$ & \textasteriskcentered  & unguided \\ 
(REASONING\_NUANCE) \textgreater  0 & $1.304$ & $0.192$ & $0.989$ & $1.619$ & $Inf$ & $1$ & \textasteriskcentered  & value guided \\ 
(REASONING\_NUANCE) \textgreater  0 & $1.453$ & $0.181$ & $1.154$ & $1.756$ & $Inf$ & $1$ & \textasteriskcentered  & controversy guided \\ 
(RESPECT) \textless  0 & $-0.095$ & $0.142$ & $-0.326$ & $0.135$ & $2.922$ & $0.745$ &  & all \\ 
(RESPECT) \textless  0 & $0.011$ & $0.212$ & $-0.334$ & $0.358$ & $0.913$ & $0.477$ &  & unguided \\ 
(RESPECT) \textless  0 & $0.010$ & $0.251$ & $-0.402$ & $0.419$ & $0.956$ & $0.489$ &  & value guided \\ 
(RESPECT) \textless  0 & $-0.364$ & $0.275$ & $-0.813$ & $0.084$ & $9.753$ & $0.907$ &  & controversy guided \\ 
(CURIOSITY) \textgreater  0 & $0.324$ & $0.089$ & $0.181$ & $0.469$ & $Inf$ & $1$ & \textasteriskcentered  & all \\ 
(CURIOSITY) \textgreater  0 & $0.508$ & $0.137$ & $0.286$ & $0.734$ & $3,999$ & $1.000$ & \textasteriskcentered  & unguided \\ 
(CURIOSITY) \textgreater  0 & $-0.011$ & $0.172$ & $-0.301$ & $0.268$ & $0.918$ & $0.478$ &  & value guided \\ 
(CURIOSITY) \textgreater  0 & $0.129$ & $0.172$ & $-0.148$ & $0.414$ & $3.386$ & $0.772$ &  & controversy guided \\ 
(COMPASSION) \textless  0 & $-0.097$ & $0.076$ & $-0.222$ & $0.025$ & $9.309$ & $0.903$ &  & all \\ 
(COMPASSION) \textless  0 & $0.018$ & $0.138$ & $-0.212$ & $0.244$ & $0.801$ & $0.445$ &  & unguided \\ 
(COMPASSION) \textless  0 & $-0.128$ & $0.138$ & $-0.350$ & $0.105$ & $4.722$ & $0.825$ &  & value guided \\ 
(COMPASSION) \textless  0 & $0.014$ & $0.149$ & $-0.230$ & $0.259$ & $0.863$ & $0.463$ &  & controversy guided \\ 
(PERSONAL\_STORY) \textless  0 & $-0.441$ & $0.092$ & $-0.595$ & $-0.293$ & $Inf$ & $1$ & \textasteriskcentered  & all \\ 
(PERSONAL\_STORY) \textless  0 & $-0.485$ & $0.161$ & $-0.749$ & $-0.216$ & $999$ & $0.999$ & \textasteriskcentered  & unguided \\ 
(PERSONAL\_STORY) \textless  0 & $-0.451$ & $0.181$ & $-0.744$ & $-0.150$ & $159$ & $0.994$ & \textasteriskcentered  & value guided \\ 
(PERSONAL\_STORY) \textless  0 & $-0.705$ & $0.180$ & $-0.998$ & $-0.411$ & $Inf$ & $1$ & \textasteriskcentered  & controversy guided \\ 
(AFFINITY) \textgreater  0 & $0.129$ & $0.143$ & $-0.103$ & $0.368$ & $4.195$ & $0.807$ &  & all \\ 
(AFFINITY) \textgreater  0 & $0.080$ & $0.253$ & $-0.333$ & $0.513$ & $1.653$ & $0.623$ &  & unguided \\ 
(AFFINITY) \textgreater  0 & $0.195$ & $0.270$ & $-0.258$ & $0.635$ & $3.292$ & $0.767$ &  & value guided \\ 
(AFFINITY) \textgreater  0 & $0.364$ & $0.261$ & $-0.059$ & $0.792$ & $11.121$ & $0.917$ &  $\dagger$ & controversy guided \\ 
(TOXICITY) \textgreater  0 & $0.729$ & $0.285$ & $0.259$ & $1.195$ & $141.857$ & $0.993$ & \textasteriskcentered  & all \\ 
(TOXICITY) \textgreater  0 & $0.225$ & $0.592$ & $-0.760$ & $1.180$ & $1.903$ & $0.655$ &  & unguided \\ 
(TOXICITY) \textgreater  0 & $0.726$ & $0.675$ & $-0.398$ & $1.817$ & $6.181$ & $0.861$ &  & value guided \\ 
(TOXICITY) \textgreater  0 & $0.914$ & $0.392$ & $0.268$ & $1.555$ & $124$ & $0.992$ & \textasteriskcentered  & controversy guided \\ 

\hline
\multicolumn{9}{c}{(b) \textbf{Analysis 1b:} Score} \\
\hline
(REASONING\_NUANCE) \textgreater  0 & $1.217$ & $0.042$ & $1.150$ & $1.286$ & $Inf$ & $1$ & \textasteriskcentered  & all \\ 
(REASONING\_NUANCE) \textgreater  0 & $1.018$ & $0.065$ & $0.911$ & $1.124$ & $Inf$ & $1$ & \textasteriskcentered  & unguided \\ 
(REASONING\_NUANCE) \textgreater  0 & $1.261$ & $0.076$ & $1.136$ & $1.385$ & $Inf$ & $1$ & \textasteriskcentered  & value guided \\ 
(REASONING\_NUANCE) \textgreater  0 & $1.571$ & $0.082$ & $1.434$ & $1.706$ & $Inf$ & $1$ & \textasteriskcentered  & controversy guided \\ 
(RESPECT) \textless  0 & $-0.057$ & $0.067$ & $-0.167$ & $0.054$ & $4.181$ & $0.807$ &  & all \\ 
(RESPECT) \textless  0 & $0.037$ & $0.112$ & $-0.151$ & $0.219$ & $0.583$ & $0.368$ &  & unguided \\ 
(RESPECT) \textless  0 & $-0.084$ & $0.113$ & $-0.270$ & $0.105$ & $3.334$ & $0.769$ &  & value guided \\ 
(RESPECT) \textless  0 & $-0.363$ & $0.137$ & $-0.588$ & $-0.135$ & $221.222$ & $0.996$ & \textasteriskcentered  & controversy guided \\ 
(CURIOSITY) \textless  0 & $-0.159$ & $0.042$ & $-0.229$ & $-0.092$ & $Inf$ & $1$ & \textasteriskcentered  & all \\ 
(CURIOSITY) \textless  0 & $-0.103$ & $0.067$ & $-0.215$ & $0.010$ & $13.493$ & $0.931$ & $\dagger$ & unguided \\ 
(CURIOSITY) \textless  0 & $-0.324$ & $0.072$ & $-0.443$ & $-0.205$ & $Inf$ & $1$ & \textasteriskcentered  & value guided \\ 
(CURIOSITY) \textless  0 & $-0.287$ & $0.086$ & $-0.428$ & $-0.145$ & $3,999$ & $1.000$ & \textasteriskcentered  & controversy guided \\ 
(COMPASSION) \textgreater  0 & $0.176$ & $0.038$ & $0.112$ & $0.239$ & $3,999$ & $1.000$ & \textasteriskcentered  & all \\ 
(COMPASSION) \textgreater  0 & $0.274$ & $0.074$ & $0.155$ & $0.399$ & $Inf$ & $1$ & \textasteriskcentered  & unguided \\ 
(COMPASSION) \textgreater  0 & $0.299$ & $0.064$ & $0.195$ & $0.406$ & $Inf$ & $1$ & \textasteriskcentered  & value guided \\ 
(COMPASSION) \textgreater  0 & $0.116$ & $0.079$ & $-0.014$ & $0.245$ & $13.545$ & $0.931$ & $\dagger$ & controversy guided \\ 
(PERSONAL\_STORY) \textless  0 & $-0.650$ & $0.044$ & $-0.723$ & $-0.577$ & $Inf$ & $1$ & \textasteriskcentered  & all \\ 
(PERSONAL\_STORY) \textless  0 & $-0.804$ & $0.080$ & $-0.934$ & $-0.672$ & $Inf$ & $1$ & \textasteriskcentered  & unguided \\ 
(PERSONAL\_STORY) \textless  0 & $-0.758$ & $0.078$ & $-0.883$ & $-0.631$ & $Inf$ & $1$ & \textasteriskcentered  & value guided \\ 
(PERSONAL\_STORY) \textless  0 & $-0.826$ & $0.088$ & $-0.973$ & $-0.682$ & $Inf$ & $1$ & \textasteriskcentered  & controversy guided \\ 
(AFFINITY) \textgreater  0 & $-0.023$ & $0.069$ & $-0.135$ & $0.089$ & $0.592$ & $0.372$ &  & all \\ 
(AFFINITY) \textgreater  0 & $0.090$ & $0.127$ & $-0.116$ & $0.300$ & $3.149$ & $0.759$ &  & unguided \\ 
(AFFINITY) \textgreater  0 & $0.055$ & $0.119$ & $-0.139$ & $0.250$ & $2.084$ & $0.676$ &  & value guided \\ 
(AFFINITY) \textgreater  0 & $0.326$ & $0.135$ & $0.107$ & $0.545$ & $141.857$ & $0.993$ & \textasteriskcentered  & controversy guided \\ 
(TOXICITY) \textless  0 & $-0.659$ & $0.123$ & $-0.865$ & $-0.458$ & $Inf$ & $1$ & \textasteriskcentered  & all \\ 
(TOXICITY) \textless  0 & $-0.123$ & $0.251$ & $-0.541$ & $0.289$ & $2.203$ & $0.688$ &  & unguided \\ 
(TOXICITY) \textless  0 & $-0.243$ & $0.263$ & $-0.679$ & $0.192$ & $4.610$ & $0.822$ &  & value guided \\ 
(TOXICITY) \textless  0 & $-0.655$ & $0.189$ & $-0.969$ & $-0.341$ & $1,332.333$ & $0.999$ & \textasteriskcentered  & controversy guided \\ 
\hline
\multicolumn{9}{c}{Interaction (dv: log-transformed score)} \\
\hline
(TOXICITY x TOXICIT\_HUMAN) \textgreater  0 &	0.244  &	0.104 &	0.078 &	0.419 &	141.857 & 0.993 &	\textasteriskcentered & all \\
\hline \\[-1.8ex] 
\end{tabular}
\end{adjustbox}
\label{tab:mainfindings}
\end{table}

\begin{table}[!htbp] \centering 
\caption{Results of hypothesis tests for moderation analysis of score (one linear regression; the score was log-transformed). In the ``Star'' column, $\ast$ indicates a BF $> 100$, our significance threshold indicating ``definitive'' or ``extreme'' evidence in favor of the alternative hypothesis, while $\dagger$ represents BF $> 10$. Note: $Inf$ should be interpreted as the maximum number of draws (12,000). BF calculated as Evidence Ratio.}
\label{tab:moderation}
\begin{adjustbox}{width=\textwidth}
\begin{tabular}{lccccccc}
\\[-1.8ex]\hline 
\hline \\[-1.8ex] 
Hypothesis & Estimate & Est.Error & CI.Lower & CI.Upper & BF & Post.Prob & Star \\ 
\hline \\[-1.8ex] 
(youngerTRUE:TOXICITY) \textless  0 & $-0.864$ & $0.268$ & $-1.301$ & $-0.419$ & $799$ & $0.999$ & \textasteriskcentered  \\ 
(maleTRUE:TOXICITY) \textgreater  0 & $0.037$ & $0.243$ & $-0.362$ & $0.442$ & $1.295$ & $0.564$ &  \\ 
(whiteTRUE:TOXICITY) \textless  0 & $-0.625$ & $0.270$ & $-1.066$ & $-0.182$ & $96.561$ & $0.990$ & $\dagger$  \\ 
(high\_valuesTRUE:TOXICITY) \textgreater  0 & $0.202$ & $0.243$ & $-0.200$ & $0.604$ & $3.916$ & $0.797$ &  \\ 
(youngerTRUE:AFFINITY) \textless  0 & $-0.030$ & $0.147$ & $-0.271$ & $0.211$ & $1.365$ & $0.577$ &  \\ 
(maleTRUE:AFFINITY) \textgreater  0 & $0.105$ & $0.139$ & $-0.123$ & $0.332$ & $3.413$ & $0.773$ &  \\ 
(whiteTRUE:AFFINITY) \textgreater  0 & $0.001$ & $0.154$ & $-0.254$ & $0.253$ & $1.020$ & $0.505$ &  \\ 
(high\_valuesTRUE:AFFINITY) \textgreater  0 & $0.235$ & $0.139$ & $0.008$ & $0.465$ & $22.033$ & $0.957$ & $\dagger$  \\ 
(youngerTRUE:COMPASSION) \textgreater  0 & $0.136$ & $0.088$ & $-0.009$ & $0.281$ & $15.438$ & $0.939$ & $\dagger$ \\ 
(maleTRUE:COMPASSION) \textless  0 & $-0.074$ & $0.080$ & $-0.207$ & $0.057$ & $4.613$ & $0.822$ &  \\ 
(whiteTRUE:COMPASSION) \textless  0 & $-0.029$ & $0.089$ & $-0.177$ & $0.119$ & $1.679$ & $0.627$ &  \\ 
(high\_valuesTRUE:COMPASSION) \textgreater  0 & $0.026$ & $0.080$ & $-0.104$ & $0.157$ & $1.670$ & $0.626$ &  \\ 
(youngerTRUE:CURIOSITY) \textgreater  0 & $0.162$ & $0.082$ & $0.028$ & $0.295$ & $37.462$ & $0.974$ & $\dagger$  \\ 
(maleTRUE:CURIOSITY) \textgreater  0 & $0.015$ & $0.075$ & $-0.108$ & $0.138$ & $1.391$ & $0.582$ &  \\ 
(whiteTRUE:CURIOSITY) \textless  0 & $-0.086$ & $0.082$ & $-0.220$ & $0.048$ & $5.889$ & $0.855$ &  \\ 
(high\_valuesTRUE:CURIOSITY) \textgreater  0 & $0.220$ & $0.077$ & $0.093$ & $0.345$ & $427.571$ & $0.998$ & \textasteriskcentered  \\ 
(youngerTRUE:REASONING\_NUANCE) \textgreater  0 & $0.088$ & $0.079$ & $-0.041$ & $0.219$ & $6.643$ & $0.869$ &  \\ 
(maleTRUE:REASONING\_NUANCE) \textgreater  0 & $0.134$ & $0.073$ & $0.015$ & $0.255$ & $29.769$ & $0.968$ & $\dagger$  \\ 
(whiteTRUE:REASONING\_NUANCE) \textgreater  0 & $0.305$ & $0.079$ & $0.175$ & $0.435$ & $11,999$ & $1.000$ & \textasteriskcentered  \\ 
(high\_valuesTRUE:REASONING\_NUANCE) \textless  0 & $-0.294$ & $0.073$ & $-0.414$ & $-0.172$ & $Inf$ & $1$ & \textasteriskcentered  \\ 
(youngerTRUE:PERSONAL\_STORY) \textless  0 & $-0.318$ & $0.096$ & $-0.477$ & $-0.159$ & $11,999$ & $1.000$ & \textasteriskcentered  \\ 
(maleTRUE:PERSONAL\_STORY) \textless  0 & $-0.104$ & $0.090$ & $-0.253$ & $0.043$ & $6.963$ & $0.874$ &  \\ 
(whiteTRUE:PERSONAL\_STORY) \textless  0 & $-0.143$ & $0.097$ & $-0.304$ & $0.014$ & $13.563$ & $0.931$ &  $\dagger$ \\ 
(high\_valuesTRUE:PERSONAL\_STORY) \textless  0 & $-0.091$ & $0.090$ & $-0.239$ & $0.059$ & $5.376$ & $0.843$ &  \\ 
(youngerTRUE:RESPECT) \textgreater  0 & $0.034$ & $0.142$ & $-0.201$ & $0.269$ & $1.474$ & $0.596$ &  \\ 
(maleTRUE:RESPECT) \textless  0 & $-0.094$ & $0.132$ & $-0.313$ & $0.120$ & $3.240$ & $0.764$ &  \\ 
(whiteTRUE:RESPECT) \textgreater  0 & $0.072$ & $0.149$ & $-0.172$ & $0.315$ & $2.146$ & $0.682$ &  \\ 
(high\_valuesTRUE:RESPECT) \textless  0 & $-0.096$ & $0.136$ & $-0.318$ & $0.128$ & $3.157$ & $0.759$ &  \\ 
\hline \\[-1.8ex] 
\end{tabular} 
\end{adjustbox}
\end{table} 

\begin{table}[!htbp] \centering 
\caption{Results of hypothesis tests for user prompt attributes predicting LLM response attribute (linear regressions; the score was log-transformed). In the ``Star'' column, $\ast$ indicates a BF $> 100$, our significance threshold indicating ``definitive'' or ``extreme'' evidence in favor of the alternative hypothesis, while $\dagger$ represents BF $> 10$. Note: $Inf$ should be interpreted as the maximum number of draws (12,000). BF calculated as Evidence Ratio.} 
\label{tab:users} 
\begin{adjustbox}{width=\textwidth}
\begin{tabular}{lcccccccc} 
\\[-1.8ex]\hline 
\hline \\[-1.8ex] 
Hypothesis & Estimate & Est.Error & CI.Lower & CI.Upper & BF & Post.Prob & Star & DV \\ 
\hline \\[-1.8ex] 
(TOXICITY\_human) \textgreater  0 & $0.197$ & $0.004$ & $0.191$ & $0.203$ & $Inf$ & $1$ & \textasteriskcentered  & TOXICITY \\ 
(AFFINITY\_human) \textgreater  0 & $0.009$ & $0.006$ & $-0.001$ & $0.018$ & $15.043$ & $0.938$ & $\dagger$ & TOXICITY \\ 
(COMPASSION\_human) \textless  0 & $-0.012$ & $0.004$ & $-0.018$ & $-0.006$ & $922.077$ & $0.999$ & \textasteriskcentered  & TOXICITY \\ 
(CURIOSITY\_human) \textgreater  0 & $0.007$ & $0.002$ & $0.004$ & $0.010$ & $11,999$ & $1.000$ & \textasteriskcentered  & TOXICITY \\ 
(REASONING\_NUANCE\_human) \textgreater  0 & $0.028$ & $0.007$ & $0.016$ & $0.040$ & $Inf$ & $1$ & \textasteriskcentered  & TOXICITY \\ 
(PERSONAL\_STORY\_human) \textless  0 & $-0.018$ & $0.003$ & $-0.023$ & $-0.013$ & $Inf$ & $1$ & \textasteriskcentered  & TOXICITY \\ 
(RESPECT\_human) \textless  0 & $-0.007$ & $0.006$ & $-0.016$ & $0.002$ & $9.545$ & $0.905$ &  & TOXICITY \\ 
(TOXICITY\_human) \textgreater  0 & $0.115$ & $0.013$ & $0.093$ & $0.137$ & $Inf$ & $1$ & \textasteriskcentered  & AFFINITY \\ 
(AFFINITY\_human) \textgreater  0 & $0.626$ & $0.022$ & $0.591$ & $0.663$ & $Inf$ & $1$ & \textasteriskcentered  & AFFINITY \\ 
(COMPASSION\_human) \textgreater  0 & $0.235$ & $0.014$ & $0.212$ & $0.258$ & $Inf$ & $1$ & \textasteriskcentered  & AFFINITY \\ 
(CURIOSITY\_human) \textless  0 & $-0.036$ & $0.007$ & $-0.047$ & $-0.024$ & $11,999$ & $1.000$ & \textasteriskcentered  & AFFINITY \\ 
(REASONING\_NUANCE\_human) \textless  0 & $-0.082$ & $0.028$ & $-0.128$ & $-0.036$ & $630.579$ & $0.998$ & \textasteriskcentered  & AFFINITY \\ 
(PERSONAL\_STORY\_human) \textless  0 & $-0.223$ & $0.011$ & $-0.242$ & $-0.205$ & $Inf$ & $1$ & \textasteriskcentered  & AFFINITY \\ 
(RESPECT\_human) \textless  0 & $-0.112$ & $0.021$ & $-0.146$ & $-0.078$ & $Inf$ & $1$ & \textasteriskcentered  & AFFINITY \\ 
(TOXICITY\_human) \textgreater  0 & $0.288$ & $0.015$ & $0.264$ & $0.312$ & $Inf$ & $1$ & \textasteriskcentered  & COMPASSION \\ 
(AFFINITY\_human) \textgreater  0 & $0.056$ & $0.024$ & $0.016$ & $0.095$ & $89.226$ & $0.989$ & $\dagger$  & COMPASSION \\ 
(COMPASSION\_human) \textgreater  0 & $1.084$ & $0.015$ & $1.059$ & $1.109$ & $Inf$ & $1$ & \textasteriskcentered  & COMPASSION \\ 
(CURIOSITY\_human) \textgreater  0 & $0.034$ & $0.008$ & $0.021$ & $0.047$ & $Inf$ & $1$ & \textasteriskcentered  & COMPASSION \\ 
(REASONING\_NUANCE\_human) \textless  0 & $-0.362$ & $0.030$ & $-0.411$ & $-0.313$ & $Inf$ & $1$ & \textasteriskcentered  & COMPASSION \\ 
(PERSONAL\_STORY\_human) \textgreater  0 & $0.045$ & $0.012$ & $0.026$ & $0.065$ & $Inf$ & $1$ & \textasteriskcentered  & COMPASSION \\ 
(RESPECT\_human) \textless  0 & $-0.458$ & $0.023$ & $-0.497$ & $-0.420$ & $Inf$ & $1$ & \textasteriskcentered  & COMPASSION \\ 
(TOXICITY\_human) \textless  0 & $-0.005$ & $0.012$ & $-0.024$ & $0.015$ & $1.859$ & $0.650$ &  & CURIOSITY \\ 
(AFFINITY\_human) \textgreater  0 & $0.164$ & $0.019$ & $0.133$ & $0.196$ & $Inf$ & $1$ & \textasteriskcentered  & CURIOSITY \\ 
(COMPASSION\_human) \textgreater  0 & $0.067$ & $0.012$ & $0.048$ & $0.088$ & $Inf$ & $1$ & \textasteriskcentered  & CURIOSITY \\ 
(CURIOSITY\_human) \textless  0 & $-0.078$ & $0.006$ & $-0.088$ & $-0.067$ & $Inf$ & $1$ & \textasteriskcentered  & CURIOSITY \\ 
(REASONING\_NUANCE\_human) \textless  0 & $-0.094$ & $0.024$ & $-0.134$ & $-0.055$ & $Inf$ & $1$ & \textasteriskcentered  & CURIOSITY \\ 
(PERSONAL\_STORY\_human) \textgreater  0 & $0.057$ & $0.010$ & $0.041$ & $0.073$ & $Inf$ & $1$ & \textasteriskcentered  & CURIOSITY \\ 
(RESPECT\_human) \textgreater  0 & $0.126$ & $0.019$ & $0.095$ & $0.156$ & $Inf$ & $1$ & \textasteriskcentered  & CURIOSITY \\ 
(TOXICITY\_human) \textgreater  0 & $0.103$ & $0.013$ & $0.081$ & $0.126$ & $Inf$ & $1$ & \textasteriskcentered  & REASONING\_NUANCE \\ 
(AFFINITY\_human) \textless  0 & $-0.101$ & $0.021$ & $-0.136$ & $-0.066$ & $Inf$ & $1$ & \textasteriskcentered  & REASONING\_NUANCE \\ 
(COMPASSION\_human) \textgreater  0 & $0.133$ & $0.014$ & $0.111$ & $0.156$ & $Inf$ & $1$ & \textasteriskcentered  & REASONING\_NUANCE \\ 
(CURIOSITY\_human) \textgreater  0 & $0.120$ & $0.007$ & $0.109$ & $0.132$ & $Inf$ & $1$ & \textasteriskcentered  & REASONING\_NUANCE \\ 
(REASONING\_NUANCE\_human) \textgreater  0 & $0.543$ & $0.028$ & $0.498$ & $0.589$ & $Inf$ & $1$ & \textasteriskcentered  & REASONING\_NUANCE \\ 
(PERSONAL\_STORY\_human) \textgreater  0 & $0.043$ & $0.011$ & $0.025$ & $0.061$ & $11,999$ & $1.000$ & \textasteriskcentered  & REASONING\_NUANCE \\ 
(RESPECT\_human) \textless  0 & $-0.195$ & $0.021$ & $-0.229$ & $-0.160$ & $Inf$ & $1$ & \textasteriskcentered  & REASONING\_NUANCE \\ 
(TOXICITY\_human) \textgreater  0 & $0.037$ & $0.013$ & $0.016$ & $0.058$ & $443.444$ & $0.998$ & \textasteriskcentered  & PERSONAL\_STORY \\ 
(AFFINITY\_human) \textless  0 & $-0.071$ & $0.021$ & $-0.105$ & $-0.037$ & $11,999$ & $1.000$ & \textasteriskcentered  & PERSONAL\_STORY \\ 
(COMPASSION\_human) \textgreater  0 & $0.079$ & $0.013$ & $0.057$ & $0.101$ & $Inf$ & $1$ & \textasteriskcentered  & PERSONAL\_STORY \\ 
(CURIOSITY\_human) \textless  0 & $-0.030$ & $0.007$ & $-0.041$ & $-0.018$ & $Inf$ & $1$ & \textasteriskcentered  & PERSONAL\_STORY \\ 
(REASONING\_NUANCE\_human) \textless  0 & $-0.426$ & $0.026$ & $-0.470$ & $-0.383$ & $Inf$ & $1$ & \textasteriskcentered  & PERSONAL\_STORY \\ 
(PERSONAL\_STORY\_human) \textgreater  0 & $0.242$ & $0.011$ & $0.225$ & $0.260$ & $Inf$ & $1$ & \textasteriskcentered  & PERSONAL\_STORY \\ 
(RESPECT\_human) \textgreater  0 & $0.329$ & $0.020$ & $0.296$ & $0.361$ & $Inf$ & $1$ & \textasteriskcentered  & PERSONAL\_STORY \\ 
(TOXICITY\_human) \textgreater  0 & $0.131$ & $0.013$ & $0.111$ & $0.152$ & $Inf$ & $1$ & \textasteriskcentered  & RESPECT \\ 
(AFFINITY\_human) \textgreater  0 & $0.197$ & $0.020$ & $0.164$ & $0.230$ & $Inf$ & $1$ & \textasteriskcentered  & RESPECT \\ 
(COMPASSION\_human) \textgreater  0 & $0.202$ & $0.013$ & $0.181$ & $0.224$ & $Inf$ & $1$ & \textasteriskcentered  & RESPECT \\ 
(CURIOSITY\_human) \textless  0 & $-0.053$ & $0.007$ & $-0.064$ & $-0.042$ & $Inf$ & $1$ & \textasteriskcentered  & RESPECT \\ 
(REASONING\_NUANCE\_human) \textless  0 & $-0.122$ & $0.026$ & $-0.164$ & $-0.080$ & $Inf$ & $1$ & \textasteriskcentered  & RESPECT \\ 
(PERSONAL\_STORY\_human) \textless  0 & $-0.121$ & $0.010$ & $-0.138$ & $-0.104$ & $Inf$ & $1$ & \textasteriskcentered  & RESPECT \\ 
(RESPECT\_human) \textgreater  0 & $0.397$ & $0.020$ & $0.365$ & $0.430$ & $Inf$ & $1$ & \textasteriskcentered  & RESPECT \\ 
\hline \\[-1.8ex] 
\end{tabular} 
\end{adjustbox}
\end{table} 

\end{document}